# DPSeq: A Novel and Efficient Digital Pathology Classifier for Predicting Cancer Biomarkers using Sequencer Architecture

Min Cen[1], Xingyu Li[2], Bangwei Guo[1], Jitendra Jonnagaddala[3*], Hong Zhang[2*], Xu Steven Xu[4*]

[1]School of Data Science, University of Science and Technology of China
[2]Department of Statistics and Finance, School of Management, University of Science and Technology of China
[3]School of Population Health, UNSW Sydney, Kensington, NSW, Australia
[4]Clinical Pharmacology and Quantitative Science, Genmab Inc., Princeton, New Jersey, USA
*Corresponding author. jitendra.jonnagaddala@unsw.edu.au (Jitendra Jonnagaddala), zhangh@ustc.edu.cn (Hong Zhang), sxu@genmab.com (Xu Steven Xu)

**Conflict of interest statements**：XSX is an employee of Genmab, Inc. Genmab did not provide any funding for this study.

**Abstract**

In digital pathology tasks, transformers have achieved state-of-the-art results, surpassing convolutional neural networks (CNNs). However, transformers are usually complex and resource intensive. In this study, we developed a novel and efficient digital pathology classifier called DPSeq, to predict cancer biomarkers through fine-tuning a sequencer architecture integrating horizon and vertical bidirectional long short-term memory (BiLSTM) networks. Using hematoxylin and eosin (H&E)-stained histopathological images of colorectal cancer (CRC) from two international datasets: The Cancer Genome Atlas (TCGA) and Molecular and Cellular Oncology (MCO), the predictive performance of DPSeq was evaluated in series of experiments. DPSeq demonstrated exceptional performance for predicting key biomarkers in CRC (MSI status, Hypermutation, CIMP status, BRAF mutation, TP53 mutation and chromosomal instability [CING]), outperforming most published state-of-the-art classifiers in a within-cohort internal validation and a cross-cohort external validation. Additionally, under the same experimental conditions using the same set of training and testing datasets, DPSeq surpassed 4 CNN (ResNet18, ResNet50, MobileNetV2, and EfficientNet) and 2 transformer (ViT and Swin-T) models, achieving the highest AUROC and AUPRC values in predicting MSI status, BRAF mutation, and CIMP status. Furthermore, DPSeq required less time for both training and prediction due to its simple architecture. Therefore, DPSeq appears to be the preferred choice over transformer and CNN models for predicting cancer biomarkers.

# Highlights

- Although emerging transformer-based digital pathology models tend to outperform CNN networks, these models are very complex and time-consuming.

- We developed a novel and efficient digital pathology classifier (DPSeq) to predict cancer biomarkers through fine-tuning a sequencer architecture integrating horizon and vertical bidirectional long short-term memory (BiLSTM) networks.

- DPSeq's predictive performance significantly surpassed that of CNN models.

- DPSeq exhibited superior predictive performance compared to advanced transformer models, while requiring less time for training and prediction due to its simpler architecture.

- Overall, DPSeq appears to be the preferred choice over transformer and CNN models for predicting cancer biomarkers.

# 1. Introduction

The advent of transformers (Vaswani et al., 2017) has ushered in a new era for Natural language processing (NLP). Drawing inspiration from this success, Dosovitskiy et al. (Dosovitskiy et al., 2020) proposed the vision transformer (ViT) for image analysis. Since then, numerous variants of the ViT have been introduced and have achieved state-of-the-art results, surpassing CNNs (Liu et al., 2021; Mehta and Rastegari, 2021; Zhai et al., 2022). In the context of digital pathology, Laleh et al. (Laleh et al., 2022) demonstrated that a ViT model outperforms certain CNN architectures in various tasks. Our recent research has shown that Swin-T transformers deliver superior predictive performance in predicting microsatellite instability (MSI) and other biomarkers in colorectal cancer and are more resilient than CNN models when dealing with limited training data (Guo et al., 2023).

Even though transformers have shown impressive results in computer vision tasks, they require significant computational resources to train on large images (He et al., 2022). Pathology images, especially whole-slide images (WSIs), are massive and contain billions of pixels, making it extremely difficult to utilize transformers for their processing. Hence, there is a pressing need to devise digital pathology models that can match or surpass the performance of transformers while being less computationally demanding and resource intensive.

The success of transformers is thought to stem from the self-attention mechanism's ability to capture long-range dependencies. While long short-term memory (LSTM) networks can also model long-term dependencies and prevent gradient vanishing through successive time steps, their use in digital pathology is relatively limited and mainly employed for segmentation tasks(BenTaieb and Hamarneh, 2019). Recently, Tatsunami et al.(Tatsunami and Taki, 2022) recently proposed Sequencer that utilizes a BiLSTM2D network in ImageNet classification tasks for regular images.

The objective of our research was to develop an efficient digital pathology classifier (DPSeq) by finetuning the BiLSTM2D network from Sequencer to achieve state-of-the-art (SOTA) predictive performance for critical molecular pathways and gene mutations (i.e., MSI, hypermutation, chromosomal instability [CING], BRAF mutation, and CpG island methylator phenotype [CIMP]) in colorectal cancer (CRC) utilizing H&E-stained WSIs. In addition, our study provides valuable insights into the comparative performance of DPSeq versus popular transformer (ViT(Dosovitskiy et al., 2020) and Swin Transformer (Swin-T)(Zhai et al., 2022)) and CNN models (ResNet18(He et al., 2016), ResNet50(He et al., 2016), MobileNetV2(Sandler et al., 2018) and EfficientNet(Tan and Le, 2019)) in predicting pathology tasks.

## 2. Methods

The DPSeq was developed to predict cancer biomarkers using whole slide images (WSIs). As illustrated in Figure 1, the classifier development process involved four steps: (a) image preprocessing and tile selection, (b) fine-tuning a pre-trained Sequencer model (trained on the ImageNet dataset) using pathology images to build the DPSeq, (c) using the DPSeq to classify tile-level biomarkers, and (d) predicting patient-level biomarkers by aggregating the tile-level predictions. To assess the performance of DPSeq, we utilized it to predict MSI, BRAF mutation, and CIMP and other key biomarkers for CRC.

### 2.1. Datasets

In this research, we acquired histopathological whole slide images (WSIs) that were stained with H&E from two datasets related to CRC. The first dataset, named MCO-CRC(Jonnagaddala et al., 2016; Ward and Hawkins, 2015), was obtained from the Molecular and Cellular Oncology (MCO) and consists of patients who underwent curative resection for CRC between 1994 and 2010 in New South Wales, Australia. This dataset is accessible through the SREDH Consortium (www.sredhconsortium.org, accessed on April 26, 2023), and MSI, BRAF mutation, and CIMP ground truth labels were available in this dataset. The second dataset, TCGA-CRC-DX (publicly available at https://portal.gdc.cancer.gov), contains formalin-fixed paraffin-embedded (FFPE) WSIs from two studies conducted by The Cancer Genome Atlas (TCGA), namely TCGA-COAD and TCGA-READ. In addition to the MSI, CIMP, and BRAF mutation that were present in the MCO-CRC dataset, the TCGA-CRC-DX dataset also had hypermutation, chromosomal instability (CING), and TP53 mutation information available.

In order to develop the tissue classifier and refine the DPSeq, we retrieved two public datasets that had been annotated by pathologists (these two datasets can be downloaded from https://zenodo.org/record/1214456#.ZC6beOxBxRE). These datasets are referred to as NCT-CRC-HE-100K and CRC-VAL-HE-7K, and comprise 100,000 and 7,180 H&E-stained tiles, respectively. Each tile measures 224x224 pixels at 0.5 microns per pixel (MPP) and has been color-normalized via Macenko's method(Macenko et al., 2009). The tiles were annotated with nine different tissue types, namely Adipose [ADI], background [BACK], debris [DEB], lymphocytes [LYM], mucus [MUC], smooth muscle [MUS], normal colon mucosa [NORM], cancer-associated stroma [STR], and colorectal adenocarcinoma epithelium [TUM].

## 2.2. Image Preprocessing and Tile Selection

We adopted the approach of Laleh et al.(Laleh et al., 2022) for processing whole slide images. All slides from the MCO-CRC and TCGA-CRC-DX datasets were segmented into nonoverlapping tiles of 512x512 pixels at 0.5 microns per pixel (MPP). The tiles were then color-normalized using Macenko's method(Macenko et al., 2009) and resized to 224x224 pixels to match the input requirement of the deep learning models. A tissue classifier, which was trained and tested on the NCT-CRC-HE-100K and CRC-VAL-HE-7K datasets, respectively(Guo et al., 2023), was employed to identify tumor tiles. Up to 500 randomly chosen tumor tiles from each patient were used as inputs for the subsequent DPSeq.

## 2.3. DPSeq

DPSeq's underlying structure was developed using the Sequencer framework (Tatsunami and Taki, 2022). The primary element of this network is the BiLSTM2D layer, which can integrate both horizontal and vertical patch data within a tile.

Defining the input of BiLSTM2D as $I \in \mathbb{R}^{H \times W \times C}$, we split the input to vertical patch sequences $\{I_{:,w,:} \in \mathbb{R}^{H \times C}\}_{w=1}^{W}$ and horizon patch sequences $\{I_{h,:,:} \in \mathbb{R}^{W \times C}\}_{h=1}^{H}$. For a fixed $w_0 \in [1, W] \cap \mathbb{Z}$, a vertical patch sequence $I_{:,w_0,:}$ was entered into a BiLSTM ($BiLSTM_{ver}$) to extract the vertical hidden feature: $H_{:,w_0,:}^{vertical} = BiLSTM_{ver}(I_{:,w_0,:})$. A series of vertical hidden features: $\{H_{:,w,:}^{vertical} \in \mathbb{R}^{H \times 2D}\}_{w=1}^{W}$ were obtained where $D$ is the dimension of hidden layer of $BiLSTM_{ver}$. The weights of $BiLSTM_{ver}$ for all vertical patch sequences were shared. Similarly, a series of horizon hidden features $\{H_{h,:,:}^{horizen} \in \mathbb{R}^{W \times 2D}\}_{h=1}^{H}$ were obtained by $H_{h,:,:}^{horizon} = BiLSTM_{hor}(I_{:,h,:})$, for $\forall h \in [1, H] \cap \mathbb{Z}$.

We concatenated all the vertical hidden features to matrix $H_{all}^{vertical} \in \mathbb{R}^{W \times H \times 2D}$ and all the horizon hidden features to matrix $H_{all}^{horizon} \in \mathbb{R}^{W \times H \times 2D}$. Then we obtained the hidden feature for $I$ by concatenating $H_{all}^{vertical}$ and $H_{all}^{horizon}$ to $H_{all} \in \mathbb{R}^{W \times H \times 4D}$. The output of the BiLSTM2D was obtain by a channel fusion with point-wise, full connection:

$$\hat{I} = PointWiseFC(H_{all}) = W_{all}H_{all} + b_{all} \in \mathbb{R}^{W \times H \times C}$$

The input tiles had dimensions of 224 pixels * 224 pixels, and we segmented them into a total of 32*32 smaller patches with a size of 7 pixels * 7 pixels. These patches were processed through a sequence of four stages of Sequencer blocks, each stage containing a different number of blocks (four, three, eight and four blocks, respectively). The Sequencer blocks replaced self-attention in transformer blocks with BiLSTM2D to improve the efficiency of the classifier. Following average pooling, we added a three-layer multilayer perceptron with ReLU and dropout layers (384-256-32) to extract molecular information from histopathological images. Finally, we added a classification layer to the top of the network for accurate classification.

In order to predict biomarkers from histopathological images, we fine-tuned our DPSeq classifier for multiclass tissue classification. We initialized the DPSeq parameters before the average pooling layer with the parameters of a Sequencer model that had been pre-trained on ImageNet. Using a fixed learning rate of 0.0001, Adam optimizer and cross-entropy loss, we trained DPSeq on NCT-CRC-HE-100K. To learn histological information without overfitting to the tissue classification problem, the model was fine-tuned for two epochs.

After fine-tuning the DPSeq, we replaced the last classifier layer with new linear layers to enable binary classification of CRC biomarkers. The biomarker classifiers were trained for up to 50 epochs with early stopping and a patience of 8, using a cosine annealing learning rate initially set to 0.0001. To handle class-imbalanced data, we employed weighted cross entropy loss. Finally, we obtained the patient-level biomarker score by averaging the tile-level scores of all tiles in the corresponding whole slide images.

**2.4. Experiment Design**

To assess the performance of the DPSeq classifier, we devised three experiments: 1) comparison with published models using internal cross-validation based on TCGA-CRC-DX dataset; 2) comparison with published models via external, cross-cohort validation using the same testing dataset (TCGA-CRC-DX); and 3) comparison of DPSeq with other backbone networks using the same training and testing datasets.

**2.4.1 Comparison using Internal Validation**

Deep-learning models have been developed using internal validation to predict key CRC biomarkers (including MSI, CIMP, hypermutation, CING, BRAF mutation, and TP53 mutation) based on the whole slide images from TCGA-CRC-DX(Bilal et al., 2021; Guo et al., 2023; Kather et al., 2020). We adopted the same four-fold cross-validation used in the literature and compared the DPSeq's prediction performance with the published state-of-the-art results. The dataset was split into four folds in the same way as Kather et al.(Kather et al., 2020) and/or Bilal et al.(Bilal et al., 2021) For each training iteration of the four-fold cross-validation, we randomly separated three folds for training and validation datasets with a ratio of 0.85:0.15, and used the remaining fold for testing.

### 2.4.2 Comparison using Cross-Cohort External Validation

To assess the generalization and robustness of DPSeq, we conducted an external validation using cross-cohort analysis. We trained DPSeq using the MCO-CRC dataset (n = 1138, 1026 and 364) to predict MSI, BRAF mutation, and CIMP, and then tested the model on the TCGA-CRC-DX dataset (n = 425, 500 and 235) that was unseen during the training. The testing subset of TCGA-CRC-DX has been commonly used as external testing dataset in literature(Echle et al., 2020; Echle et al., 2022; Guo et al., 2023; Laleh et al., 2022) and can provide relatively fair comparison of performance among different published models through the external cross-cohort validation. We compared DPSeq with seven published models, including Swin-T (trained on MCO-CRC [n = 1065 for MSI status,1026 for BRAF mutation]), EfficientNet (trained on DACHs [n = 2069 for MSI status and BRAF mutation ]), ViT (trained on DACHs [n = 2069 for MSI status and BRAF mutation]), ResNet18 (trained on pooled international datasets [n = 7917 for MSI status]), and ShuffleNet (trained on QUASAR [n = 1016 for MSI status], DACHs [n = 2013 for MSI status], and NLCS [n = 2197 for MSI status]).

### 2.4.3 Comparison of DPSeq with CNN networks and transformers

To avoid potential bias in comparing models and networks due to different training sets used in published articles(Echle et al., 2020; Echle et al., 2022; Guo et al., 2023; Laleh et al., 2022), we conducted an external cross-cohort validation using the same training and testing datasets. Specifically, we trained four popular CNN models (ResNet18, ResNet50, MobileNetV2, and EfficientNet) and two transformers (ViT and Swin-T) for prediction of MSI, CIMP, and BRAF mutation using the MCO-CRC dataset and evaluated their external predictive performance using the same TCGA-CRC-DX dataset. In addition to comparing the model predictive performance, we also analyzed the model efficiency by measuring training time per epoch and prediction time for MSI status across all patients in TCGA-CRC-DX.

## 2.5. Statistical analyses

To assess the predictive performance of the models, we computed the area under the receiver operating characteristic curve (AUROC) and the area under the precision-recall curve (AUPRC). For the four-fold cross-validation experiment, we obtained the average AUROC and AUPRC values across the four test folds and calculated their standard deviation. For the external validation experiment, we computed the AUROC and AUPRC values for the TCGA-CRC-DX dataset and estimated their 95% confidence intervals (CI) using the bootstrap method (1,000 iterations).

## 3. Results

### 3.1. Comparison with published models using four-fold cross-validation

DPSeq was utilized to predict six clinically relevant biomarkers for CRC, namely MSI, CIMP, hypermutation, BRAF mutation, TP53 mutation, and CING, through four-fold cross-validation on TCGA-CRC-DX dataset. To facilitate the comparison of the performance of DPSeq with previous published models, we adhered to the same TCGA-CRC-DX split as the earlier publications (Bilal et al., 2021; Guo et al., 2023; Kather et al., 2020). Table 1 presents the mean AUROC/AUPRC values from the four-fold cross-validation along with their corresponding standard deviations.

DPSeq outperformed most reported state-of-the-art models under the same experimental conditions (Table 1). DPSeq achieved the highest AUROC values of 92% (±3%) and AUPRC values of 68% (±12%) for MSI status prediction, surpassing the results reported in three recent publications(Bilal et al., 2021; Guo et al., 2023; Kather et al., 2020). Moreover, DPSeq exhibited significant improvements over the published models in predicting hypermutation and CIMP status. For hypermutation prediction, DPSeq's average AUROC and AUPRC values of 88% (±3%) and 65% (±5%) respectively, were 3-8% higher than the results reported in Guo et al.(Guo et al., 2023), Bilal et al.(Bilal et al., 2021), and Kather et al.(Kather et al., 2020), respectively. In CIMP status prediction, DPSeq achieved AUROC and AUPRC values of 81% (±4%) and 65% (±4%), respectively, which were 4-14% higher than the results reported in previous publications. Furthermore, DPSeq's predictive performance for BRAF mutation, TP53 mutation, and CING status was also competitive with state-of-the-art results. Specifically, DPSeq achieved an AUPRC value of 38% (±3%) for BRAF mutation prediction, which was 3-5% higher than the results reported in Guo et al.(Guo et al., 2023) and Bilal et al.(Bilal et al., 2021) respectively.

**3.2. Comparison with published models using cross-cohort external validation**

DPSeq was trained using the MCO-CRC dataset to predict MSI Status, BRAF Mutation, and CIMP Status. The robustness of DPSeq's predictive performance was tested using TCGA-CRC-DX datasets for these 3 biomarkers. DPSeq demonstrated exceptional performance in external validation on TCGA-CRC-DX datasets (Figure 2). DPSeq's performance in predicting MSI status resulted in an AUROC value of 89% (95% CI: 83%-94%) and an AUPRC value of 71% (95% CI: 60%-83%). When predicting BRAF mutation on TCGA-CRC, DPSeq achieved an AUROC value of 83% (95% CI: 77%-88%) and an AUPRC value of 46% (95% CI: 34%-61%). Regarding CIMP status, DPSeq demonstrated an AUROC value of 80% (95% CI: 72%-87%) and AUPRC value of 63% (95% CI: 51%-75%).

The same TCGA-CRC-DX datasets were used in the published articles for external validation of these biomarkers (Echle et al., 2020; Echle et al., 2022; Guo et al., 2023; Laleh et al., 2022), which allows fair comparison of DPSeq's predictive performance with published models (Table 2). However, it is worth noting that more training data can usually improve the predictive performance in external validation datasets. For predicting BRAF mutation and CIMP status, DPSeq significantly outperforms all other methods, even when trained with small subsets of MCO-CRC (n=1026 for BRAF mutation, n=364 for CIMP Status). DPSeq exhibited superior performance in predicting MSI status compared to ViT and EfficientNet and achieved similar results to current state-of-the-art models, including our recent Swin-T model (Guo et al., 2023) and the Resnet18 model developed by Echle et al. (Echle et al., 2022). It's worth noting that the Resnet18 model was trained on a significantly larger, multicenter dataset (N = 7917), which is almost eight times larger than the MCO-CRC dataset (N = 1138).

### 3.3. Comparison with CNN and transformer models

To ensure a fair comparison, we trained four popular CNN models (ResNet18, ResNet50, MobileNetV2, and EfficientNet) and two advanced transformer models (Vision Transformer and Swin-T) in the same way as DPSeq using the MCO-CRC dataset. We then compared their external predictive performance using the TCGA-CRC-DX dataset. As depicted in Figure 3, DPSeq surpassed all other CNN and transformer models, achieving the highest AUROC and AUPRC values in all prediction tasks (MSI status, BRAF mutation, and CIMP status). In predicting MSI status, DPSeq achieved an AUROC value of 89% (95% CI: 83%-94%) and an AUPRC value of 71% (95% CI: 60%-83%), which was about 3% higher than the transformer models and over 10% - 28% higher than the CNN models. Similarly, DPSeq's AUROC and AUPRC values in predicting BRAF mutation (AUROC = 83%; 95% CI: 77%-88% and AUPRC = 46%; 95% CI: 34%-61%) were at least 5% and 2% higher than ViT and Swin-T, respectively. Notably, DPSeq's AUPRC value was about 9% higher than that of ViT (AUPRC = 37%; 95% CI: 26%-52%). In predicting CIMP status, DPSeq slightly outperformed Swin-T but significantly outperformed other models in terms of the AUROC results. Moreover, DPSeq's AUPRC value (63%; 95%CI: 51%-75%) for predicting CIMP was approximately 8% and >20% higher than ViT and the CNN models, respectively.

### 3.4. Model complexity and time efficiency

We compared DPSeq with reference networks not only in terms of predictive performance, but also in terms of model complexity and time efficiency. As expected, larger models generally require longer training and prediction times (as shown in Figure 4 X). Despite being larger than CNN models, DPSeq is smaller than transformers and requires less training and prediction time. Moreover, DPSeq achieves superior predictive performance compared to transformer models. In contrast, although CNN-based models are much smaller and faster, their predictive performance is significantly inferior to that of DPSeq. Therefore, taking into account all three factors of model complexity, time efficiency, and predictive performance, DPSeq appears to be the preferred choice for practical applications.

## 4. Discussion

Convolutional neural networks (CNNs) (e.g., ResNet(He et al., 2016), MobileNetV2(Sandler et al., 2018) and EfficientNet(Tan and Le, 2019), etc.) have become the dominant architecture in digital pathology, including tasks such as tumor detection(Campanella et al., 2019; Pinckaers et al., 2021), subtyping(Lu et al., 2021; Wang et al., 2019; Zhu et al., 2021), and grading(Bulten et al., 2020; Shaban et al., 2020; Ström et al., 2020), and predicting molecular biomarkers using H&E-stained histopathological images. More recently, vision transformers have emerged and surpassed CNNs. However, transformers are usually extremely complex and resource-demanding due to their large model size and number of parameters.

In this study, we developed a novel and efficient digital pathology classifier called DPSeq to predict cancer biomarkers through fine-tuning a sequencer architecture integrating horizon and vertical BiLSTM networks. Based on H&E-stained histopathological images, DPSeq demonstrated exceptional performance for predicting key biomarkers in CRC (MSI status, Hypermutation, CIMP status, BRAF mutation, TP53 mutation and CING), outperforming most published state-of-the-art models in a within-cohort internal validation and a cross-cohort external validation. Additionally, under the same experimental conditions using the same set of training and testing datasets, DPSeq surpassed 4 CNN (ResNet18, ResNet50, MobileNetV2, and EfficientNet) and 2 transformer (ViT and Swin-T) models, achieving the highest AUROC and AUPRC values in predicting MSI status, BRAF mutation, and CIMP status.

Compared to current CNN and transformer models, DPSeq reduced the model size to a level comparable to that of ResNet50, while providing better or similar prediction performance than the larger and more complex transformer models such as ViT and Swin-T. Our experiments showed that DPSeq required less time for training and prediction than transformer models. Overall, DPSeq demonstrated the highest performance/complexity ratio among all CNN- and transformer-based models tested. Therefore, DPSeq appears to be the preferred choice over transformer and CNN models for predicting cancer biomarkers. The advantages of BiLSTM networks indicate that it could be a promising and practical backbone for digital pathology tasks. As such, additional research and innovation on BiLSTM architectures should be pursued in the areas of computer vision and digital pathology.

## Data Availability

The Cancer Genome Atlas is publicly available at https://portal.gdc.cancer.gov. The MCO dataset is available through the SREDH Consortium (https//: www.sredhconsortium.org,).

## Acknowledgements

The work of MC, XL, BG and HZ was partially supported by National Natural Science Foundation of China (No. 12171451 and No. 72091212) and Anhui Center for Applied Mathematics. JJ is funded by the Australian National Health and Medical Research Council (No. GNT1192469) who also acknowledges the funding support received through the Research Technology Services at UNSW Sydney, Google Cloud Research (award# GCP19980904) and NVIDIA Academic Hardware grant programs.

## Author Information

Authors and Affiliations

**School of Data Science, University of Science and Technology of China**

Min Cen, Bangwei Guo

**Department of Statistics and Finance, School of Management, University of Science and Technology of China**

Xingyu Li, Hong Zhang

**School of Population Health, UNSW Sydney, Kensington, NSW, Australia**

Jitendra Jonnagaddala

**Clinical Pharmacology and Quantitative Science, Genmab Inc., Princeton, New Jersey, USA**

Xu Steven Xu

**Corresponding authors**

Correspondence to Xu Steven Xu, Hong Zhang or Jitendra Jonnagaddala

# Ethics declarations

Access to the TCGA and MCO dataset was provided through data usage and ethics agreements.

All patients in the TCGA and MCO datasets provided written informed consent.

# Competing interests

The authors declare no competing interests.

**Table 1: Predictive performance statistics (AUROC/AUPRC) in the four-fold cross validation using TCGA-CRC-DX dataset.** Patient-level AUROC and AUPRC with their standard deviation (±SD) of four-fold cross validation of six key biomarkers or categories (MSI status, Hypermutation, CIMP status, BRAF mutation, TP53 mutation and CING) by DPSeq and other published models. The best results are in bold.

| Method | MSI Status | Hypermutation | CIMP Status | BRAF Mutation | TP53 Mutation | CING |
|---|---|---|---|---|---|---|
| AUROC (±SD) | | | | | | |
| Kather et al.(Kather et al., 2020) | 74% | 71% | -- | 66% | 64% | 73% |
| Bilal et al.(Bilal et al., 2021) | 86%±3% | 81%±4% | 79%±5% | **79%**±1% | **73%**±2% | **83**±2% |
| Guo et al.(Guo et al., 2023) | 91%±2% | 85%±3% | 77%±6% | 77%±2% | **73%**±2% | 82±4% |
| **DPSeq (Ours)** | **92%**±3% | **88%**±3% | **81%**±4% | 78%±4% | **73%**±3% | 81%±1% |
| AUPRC (±SD) | | | | | | |
| Bilal et al.(Bilal et al., 2021) | 62%±10% | 57%±9% | 51%±5% | 33%±5% | **78%**±4% | **92%**±1% |
| Guo(Guo et al., 2023) | 66%±9% | 58%±5% | 60%±15% | 35%±11% | 75%±5% | 90%±3% |
| **DPSeq (Ours)** | **68%**±12% | **65%**±5% | **65%**±4% | **38%**±3% | **78%**±5% | **92%**±2% |

**Table 2: Predictive performance statistics (AUROC/AUPRC) in the cross-cohort external validation using TCGA-CRC-DX dataset.** Patient-level AUROC and AUPRC with a 95% confidence interval obtained via bootstrapping (1,000×) calculated DPSeq and other published models in predicting MSI, BRAF mutation, and CIMP. Best results are in bold. The second column contains the dataset for training, and their number of training samples for three biomarkers. '--' means that the previous papers did not do or did not inform.

| Method | Dataset for Training | AUROC (MSI Status) | AUROC (BRAF Mutation) | AUROC (CIMP Status) | AUPRC (MSI Status) | AUPRC (BRAF Mutation) | AUPRC (CIMP Status) |
|---|---|---|---|---|---|---|---|
| ShuffleNet(Echle et al., 2020) | QUASAR N=1016, --, -- | 76% (70%-79%) | -- | -- | -- | -- | -- |
| ShuffleNet(Echle et al., 2020) | DACHs N=2013, --, -- | 77% (73%-79%) | -- | -- | -- | -- | -- |
| ShuffleNet(Echle et al., 2020) | NLCS N=2197, --, -- | 72% (71%-78%) | -- | -- | -- | -- | -- |
| EfficientNet(Laleh et al., 2022) | DACHs N=2069, 2069, -- | 88% (83%-93%) | 81% (75%-86%) | -- | 54% (44%-63%) | 36% (25%-49%) | -- |
| ViT(Laleh et al., 2022) | DACHs N=2069, 2069, -- | 89% (84%-93%) | 79% (72%-84%) | -- | 67% (56%-7%) | 30% (22%-41%) | -- |
| ResNet18(Echle et al., 2022) | Pooled Datasets N=7917 | **91% (87%-95%)** | -- | -- | -- | -- | -- |
| Swin-T(Guo et al., 2023) | MCO N=1065, 1026, -- | 90% (85%-95%) | 80% (73%-86%) | 76% (68%-84%) | **72% (61%-82%)** | 39% (28%-54%) | -- |
| DPSeq (Ours) | MCO N=1138, 1026, 364 | 89% (83%-94%) | **83% (77%-88%)** | **80% (72%-87%)** | 71% (59%-83%) | **46% (34%-61%)** | **63% (51%-75%)** |

# Figures

**Figure 1: Pipelines with DPSeq for predicting molecular pathways and gene mutations in CRC.** MCO-CRC and TCGA-CRC-DX were used to train and test for prediction of molecular biomarkers in CRC (i.e., MSI, *BRAF* mutation, and CIMP). The whole-slide images were tessellated into non-overlapping tiles of $512 \times 512$ pixels at a resolution of 0.5 µm. The resulting tiles were then resized to $224 \times 224$ pixels and color normalized. Tumor tissues (tiles) were subsequently selected by a Swin-T-based tissue-type classifier. Up to 500 tumor tiles were randomly selected for each slide. DPSeq fine-tuned by tissue classification task were trained to predict tile-level biomarkers. The predictive slide labels were obtained via tile score aggregation. At the bottom right of the figure is the core structure of BiLSTM2D in our DPSeq.

**Figure 2: ROC curves and PR curves of external validation for microsatellite for MSI, BRAF mutation and CIMP prediction on the TCGA-CRC-DX cohort.** Receiver operating characteristic curves (ROCs) and Precision recall curves (PRs) are computed for prediction of MSI, BRAF mutation and CIMP. Red-shaded areas represent the 95% confidence interval (CI) calculated via bootstrapping (1,000×). Values in the lower right of each plot indicate mean area under the receiver operating characteristic curve (AUROC; 95% CI) and the mean area under the precision-recall curve (AUPRC; 95% CI).

**Figure 3: Predictive comparison between DPSeq and other CNN-based and transformer-based models in the external validation for MSI, BRAF mutation and CIMP prediction on the TCGA-CRC-DX cohort.** Lollipop charts of AUROC and AUPRC values of DPSeq and other CNN-based (ResNet18, ResNet50, MobileNetV2, and EfficientNet) and transformer-based (ViT and Swin-T) models.

**Figure 4: Comparison of efficiency of DPSeq and other CNN-based and transformer-based models in in the external validation for MSI prediction on the TCGA-CRC-DX cohort.** Training and prediction times are plotted against number of parameters. The training time for an epoch (training set = MCO-CRC) and prediction time (for all patients in the TCGA-CRC-DX dataset) for the MSI status are recorded. Color of symbols represents the type of model (CNN, transformer, BiLSTM). Different size of symbols in the top subfigure represents number of parameters.

**Figure 1.**

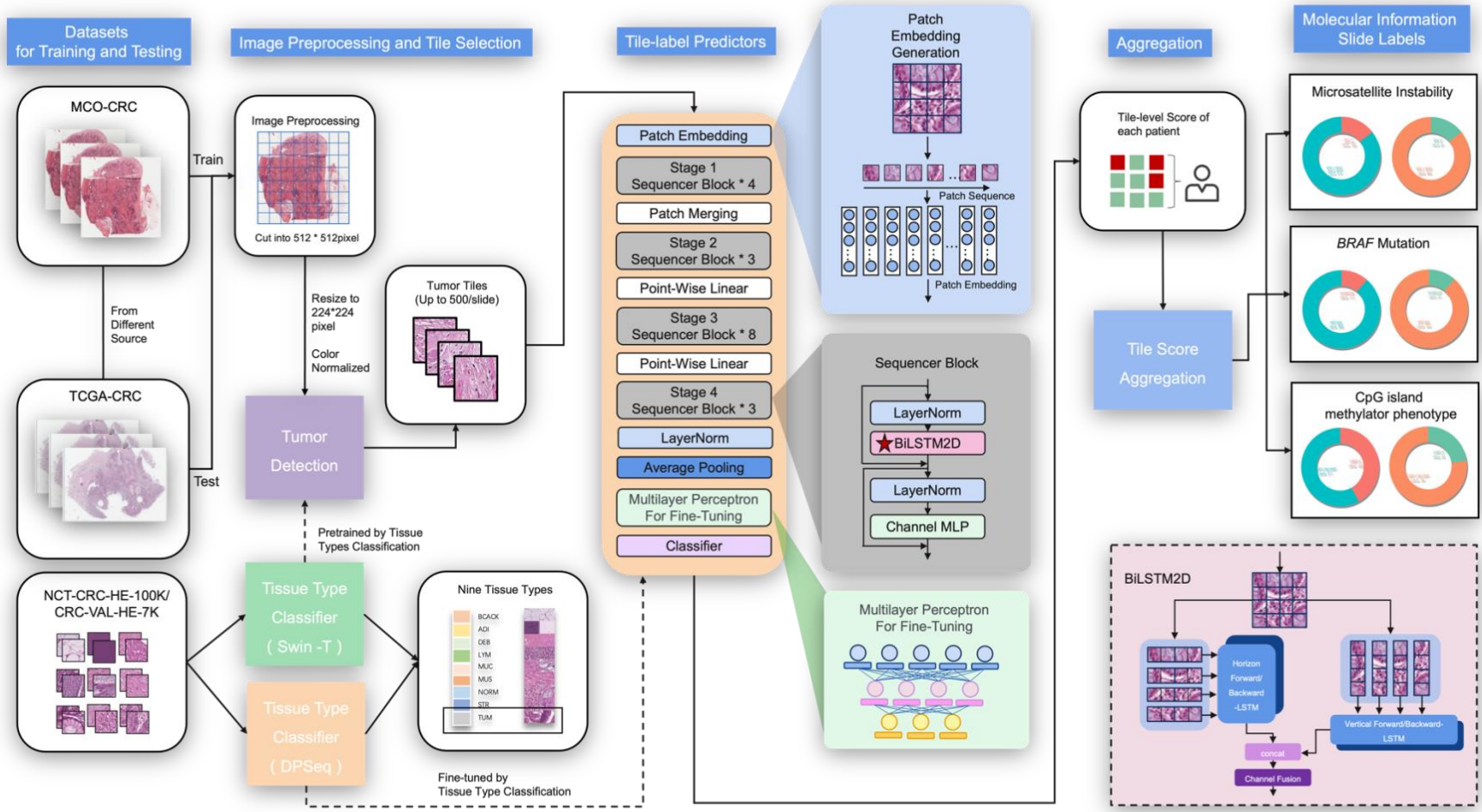

**Figure 2.**

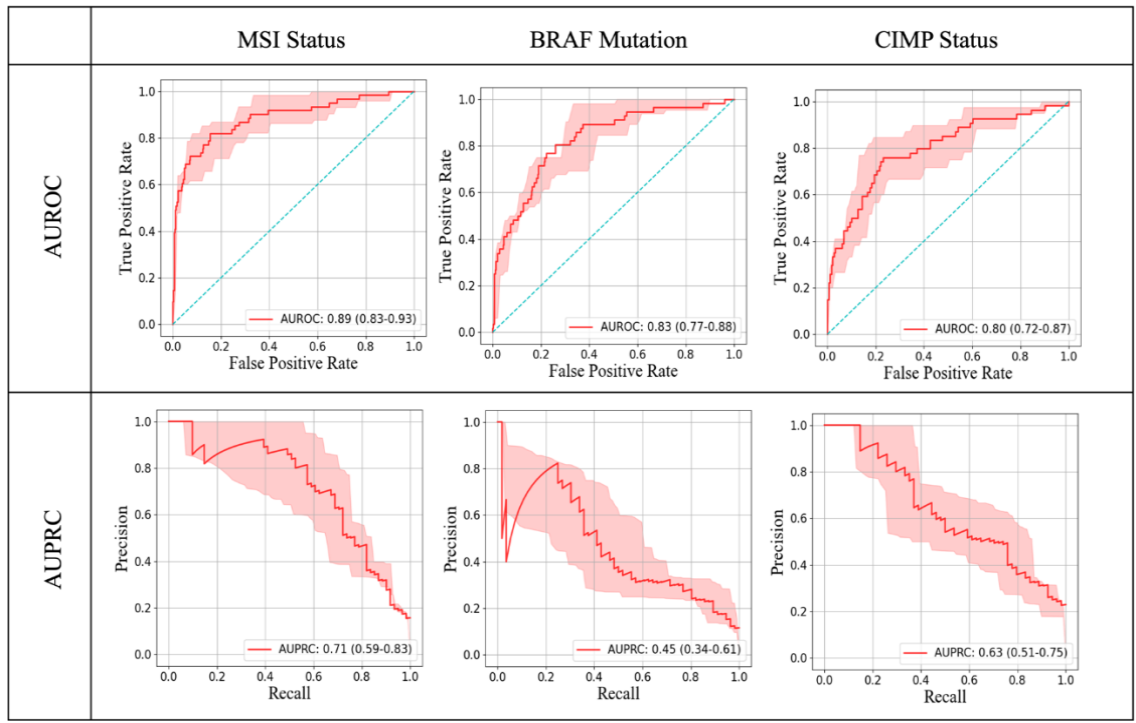

**Figure 3.**

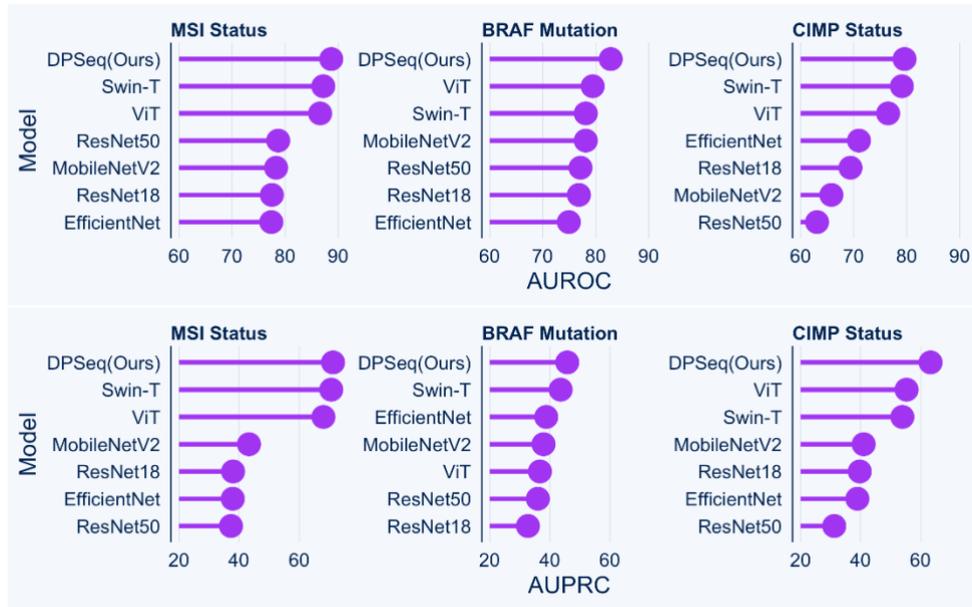

**Figure 4.**

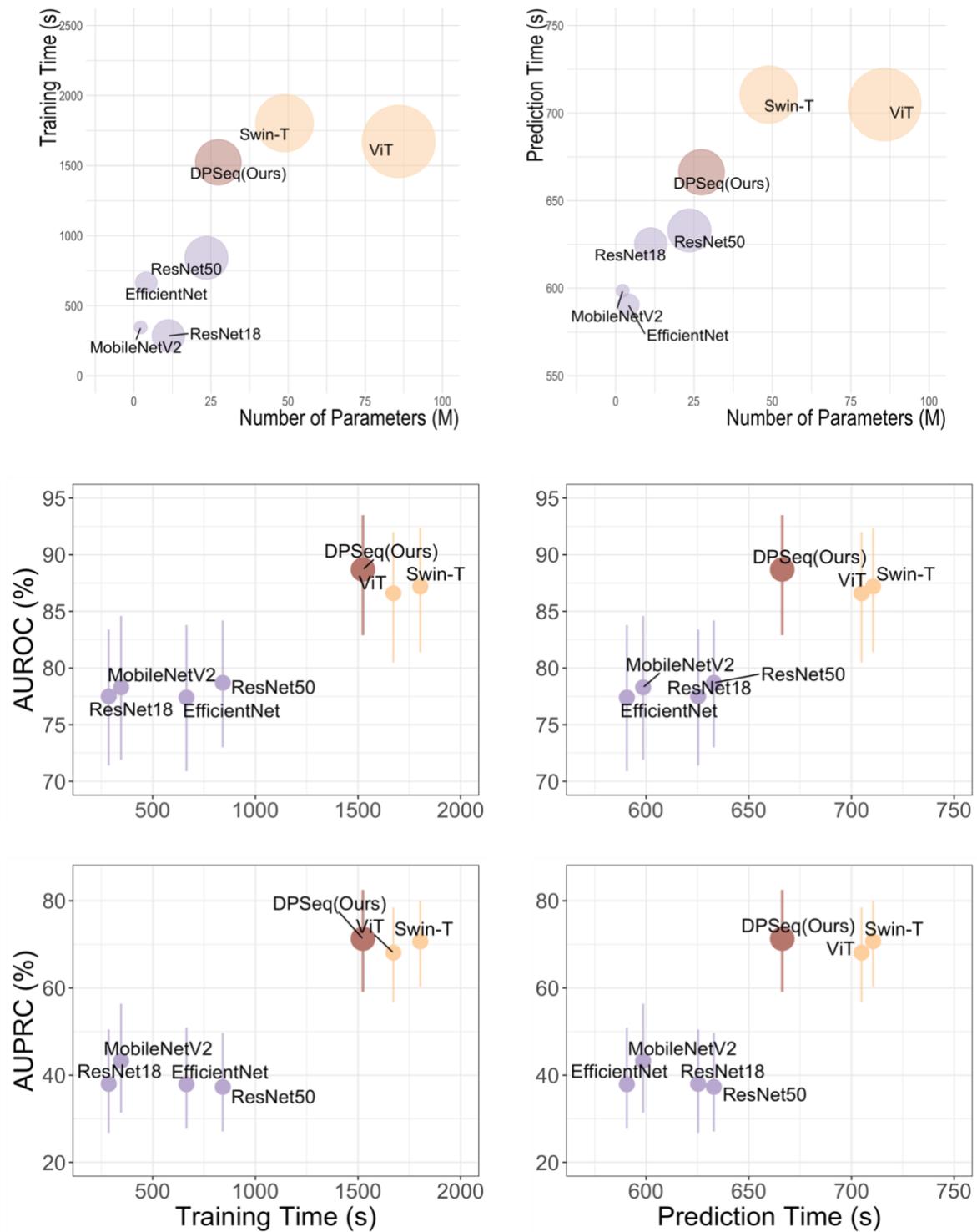

# DPSeq: A Novel and Efficient Digital Pathology Classifier for Predicting Cancer Biomarkers using Sequencer Architecture

## Supplementary Tables

**Table S1: Number of entire side images of CRC patients regarding molecular-level information prediction.** For all biomarkers (microsatellite instability [MSI], *BRAF* mutation, and CpG island methylator phenotype [CIMP]), the number of entire-slide images in Molecular and Cellular Oncology (MCO)-CRC (for training) and The Cancer Genome Atlas (TCGA)-CRC (for testing) are listed.

| Slide Label | MCO-CRC (training) | TCGA-CRC (testing) |
|---|---|---|
| Microsatellite instability (MSI-H vs MSI-L/MSS) | 1138 (166:972) | 425 (61:364) |
| BRAF mutation (Mutational vs Wild type) | 1026 (117:909) | 500 (57:443) |
| CpG island methylator phenotype (CIMP-H vs CIMP-L/None CRC CIMP) | 364 (153:211) | 235 (54:181) |